\documentclass[twocolumn,aps,showpacs,prc]{revtex4}
\usepackage[dvips]{epsfig}
\begin{document}

\title{ Radiative neutron capture reaction rates for r-process nucleosynthesis }


\author{ Vinay Singh$^{1}$, Joydev Lahiri$^{2}$, Malay Kanti Dey$^{3}$ and D. N. Basu$^{4}$}

\affiliation{Variable Energy Cyclotron Centre, 1/AF Bidhan Nagar, Kolkata 700064, INDIA}

\email[E-mail 1: ]{vsingh@vecc.gov.in}
\email[E-mail 2: ]{joy@vecc.gov.in}
\email[E-mail 3: ]{malay@vecc.gov.in}
\email[E-mail 4: ]{dnb@vecc.gov.in} 

\date{\today }

\begin{abstract}

    About half of the elements beyond iron are synthesized in stars by rapid-neutron capture process (r-process). The stellar environment provides very high neutron flux in a short time ($\sim$ seconds) which is conducive for the creation of progressively neutron-rich nuclei till the waiting point is reached after which no further neutron capture reactions proceed. At this point such extremely neutron-rich nuclei become stable via $\beta^-$ decay. A detailed understanding of the r-process remains illusive. In the present work, we explore the radiative neutron-capture (n,$\gamma$) cross sections and reaction rates around the r-process peak near mass number eighty. The inherent uncertainties remain large in some cases, particularly in case of neutron-rich nuclei. When the low-energy enhancement exists, it results in significant increase in the reaction rate for neutron-capture.
\end{abstract}

\maketitle

\noindent
\section{Introduction}
\label{section1}

    The nucleosynthesis via rapid neutron capture process (r-process) occurring in astrophysical environment is regarded as carrying the load of producing nearly half of the heavy elements abundant in nature. This broad-spectrum problem poses a great challenge for physics. The properties of nuclei like masses, neutron capture and $\beta$-decay rates along with $\beta$-delayed neutron emission probabilities are crucial physical quantities which are required for the calculations of r-process nucleosynthesis. While nuclear properties in the neighborhood of stability are well known, much remains to be explored in the domain of neutron-rich nuclei far away from the stability line which may aid the r-process. The studies of sensitivity ascertain the response of a change in nuclear physics input(s). It facilitates fixing of critical nuclear properties that decide observed final elemental abundances prevalent in nature.

    The recognition of locations responsible for the synthesis of heavy elements \cite{DC03,DC13} is a matter of great interest in astrophysics. It is well known for decades that the nuclear abundances in our solar system of elements which are heavier than iron can be partitioned in two major processes that synthesize them. In the slow neutron capture process (s-process), substantial time elapses for the $\beta$-decay to follow before capturing another neutron. This process produces nuclei around the line of stability. On the contrary, rapid neutron capture process (r-process) is rather fast in timescale producing nuclei on the neutron-rich side of the valley of stability. There also exists another process called the proton capture process (p-process) that produces primarily elements with low abundances on the proton-rich side \cite{Bu57,Ca57}. While the basic physics \cite{Wa97} of the s-process \cite{Ka11} and heavy p-process \cite{Ar03,Th10} is better understood and astrophysical locations well identified, the same do not hold in case of the r-process \cite{Qi07,Ar07,Th11}.
		
    In principle, by subtracting the p- and s-process contributions \cite{An89,Ar99} from the solar system abundances, the r-process pattern can be extracted which consists of three predominant peaks in the abundances at mass numbers $\sim$ 80, 130 and 195, linked with the neutron magic numbers at 50, 82 and 126. Of the order of 100 neutrons per seed nucleus are necessary to produce the heaviest r-process elements. Additional constraints arise from observations of r-process elements in old stars in the galactic halo \cite{Sn08,Ro14} and meteoritic data \cite{Wa96}. These data point towards specific origins for the r-process nuclei of light mass number $<$ 120 and heavy mass number $>$ 120. The distribution of main r-process elemental abundances is much the same among r-process enhanced halo stars and is similar to the residuals of the solar system. This is indicative of the fact that since early galactic times \cite{Ma90,Ar04,Ko14,Ma15} identical process created these elements that operates in a consistent manner. Though the timescale argument is obeyed by the core-collapse supernovae, the promise of early studies \cite{Me92,Wo94} is yet to be achieved by the simulations \cite{Arc07,Fi10,Hu10,Ro12} of recent times. The neutron star mergers cause low temperature outflows which are highly neutron-rich \cite{La74,Me89,Fr99,Wa14,Ju15} and due to fission recycling \cite{Go11,Ko12} have consistent abundance pattern. But mergers suffer from time delay \cite{Wa15} uncertainty. Other major astrophysical sites explored comprise hot accretion disk outflows from the  mergers of neutron stars or neutron star-black hole \cite{Wa14,Ju15,Su08,Pe14}, supernova neutron-rich jets \cite{Wi12,No15,Ts15}, gamma-ray burst collapsar outflows \cite{Pr03,Su06,Ma12}, shocked surface layers of O-Ne-Mg cores \cite{Wa03,Ja08} and neutrino-induced nucleosynthesis in the helium shell of exploding massive stars \cite{Ba11}. 
		
    The above mentioned astrophysical objects are governed by definitive environments such as initial composition, temporal distribution of density and temperature and neutron-richness which lead to elemental abundances that are unique and bear the signatures of these conditions. The preciseness and near-universality \cite{Ro10} of the pattern should have led to the main astrophysical sites for the  r-process. But the uncertainties of the nuclear rates used in the network calculations for predicting the r-process estimates would cause large scale correlations to be unreliable. Moreover, the complications involved in simulating astrophysical environments and idiosyncrasies of the properties of a large number of neutron-rich nuclei participating in the r-process lead to additional uncertainties. In the present work, the radiative neutron-capture (n,$\gamma$) cross sections and reaction rates around the r-process peak near mass number eighty have been calculated and inherent uncertainties in case of neutron-rich nuclei have been investigated.
		
\noindent
\section{Theoretical formalism}
\label{section2}

  The thermonuclear reaction rates can be obtained by convoluting fusion cross sections with Maxwell-Boltzmann distribution of energies. These cross sections can vary by several orders of magnitude across the required energy range. The low energy fusion cross sections $\sigma$, some of which are not sufficiently well known, can be obtained from laboratory experiments. However, there are cases, in particular involving the weak interaction such as the basic p$+$ p fusion to deuterium in the solar p-p chain, where no experimental data are available and one completely relies on theoretical calculations \cite{Ad11}. The theoretical estimates of the thermonuclear reaction rates depend on the various approximations used. Several factors influence the measured values of the cross sections. We need to account for the Maxwellian-averaged thermonuclear reaction rates in the network calculations used in primordial and stellar nucleosynthesis.

    The reaction rates used in the Big Bang Nucleosynthesis (BBN) reaction network have temperature dependences except $^6$Li(n,$\gamma$)$^7$Li, $^{10}$B(n,$\gamma$)$^{11}$B, $^{12}$C(n,$\gamma$)$^{13}$C and $^{14}$N(n,$\gamma$)$^{15}$N which are constant with respect to temperature. The computer code TALYS \cite{Talys} allows a comprehensive astrophysical reaction rate calculations apart from other nuclear physics calculations. To a good approximation, in the interior of stars the assumption of a thermodynamic equilibrium holds and nuclei exist both in the ground and excited states. This assumption along with cross sections calculated from compound nucleus model for various excited states facilitates Maxwellian-averaged reaction rates. For stellar evolution models this is quite an important input. The nuclear reaction rates are generally evaluated using the statistical model \cite{Rauscher97,Rauscher10} and astrophysical calculations mostly use these reaction rates. Stellar reaction rate calculations have been routinely done in past \cite{Rauscher00,Rauscher01}. However, TALYS has extended these Hauser-Feshbach statistical model (HF) \cite{Ha52} calculations by adding some new and important features. Apart from coherent inclusion of fission channel it also includes reaction mechanism that occurs before equilibrium is reached, multi-particle emission, competition among all open channels, width fluctuation corrections in detail, coupled channel description in case of deformed nuclei and level densities that are parity-dependent. The nuclear models are also normalized for available experimental data using separate approaches such as on photo-absorption data, the E1 resonance strength or on s-wave spacings, the level densities.

\subsection{ Cross sections for radiative capture }

    In the low energy domain, compound nucleus is formed by the fusion of the projectile and the target nuclei. While the total energy $E^{tot}$ is fixed from energy conservation, the total spin $J$ and parity $\Pi$ can have a range of values. The reaction obeys the following conservation laws,
$$
E_{a}+S_{a}\ =\ E_{a'}+E_{x}+S_{a'}=E^{tot},~~~~{\rm energy~conservation},  
$$
$$
s+I+l\ =\ s'+I'+l'=J,~~~~{\rm angular~momentum~conservation},
$$
\begin{center}
  $\pi_{0}\Pi_{0}(-1)^{l}\ = \pi_f\Pi_f(-1)^{l'}=\Pi,~~~~{\rm parity~conservation}.$
\end{center}
The formula for binary cross section, assuming the compound nucleus model, is given by

\begin{eqnarray}
\sigma_{\alpha\alpha'}^{comp}\ &&=\ D^{comp}\frac{\pi}{k^{2}}\sum_{J=mod(I+s,1)}^{l_{\max}+I+s}
\sum_{\Pi=-1}^{1}\frac{2J+1}{(2I+1)(2s+1)}  \nonumber\\
&&\sum_{j=|J-I|}^{J+I}\sum_{l=|j-s|}^{j+s}\sum_{j'=|J-I'|}^{J+I'}\sum_{l'=|j'-s'|}^{j'+s'} \delta_{\pi}(\alpha)\delta_{\pi}(\alpha')
\end{eqnarray}

\begin{center}
   $\displaystyle \times\ \frac{T_{\alpha lj}^{J}(E_{a})\langle T_{\alpha' l'j'}^{J}(E_{a'})\rangle}{\sum_{\alpha'',l'',j''}\delta_{\pi}(\alpha'')\langle T_{\alpha''l''j''}^{J}(E_{a''})\rangle}W_{\alpha lj\alpha'l'j'}^{J}$
\end{center}
\noindent
where

$E_{a}=$ the energy of the projectile 

$l=$ the orbital angular momentum of the projectile

$s=$ the spin of the projectile

$j=$ the total angular momentum of the projectile

$\pi_{0}=$ the parity of the projectile

$\delta_{\pi}(\alpha)=\left\{ \begin{array}{ll}
1 ~~&{\rm if}~~(-1)^{l}\pi_{0}\Pi_{0}=\Pi \\ 
 0~~& {\rm otherwise}
\end{array}\right.$ 

$\alpha=$ the designation of the channel for the initial projectile-target system:

$\alpha=\{a,\ s,\ E_{a},\ E_{x}^{0},\ I,\ \Pi_{0}\}$, where $a$ and $E_{x}^{0}$ are the type of the projectile and  the excitation energy (which is zero usually) of the target nucleus, respectively 

$l_{\max}=$ the maximum l-value of the projectile

$S_{a}=$ the separation energy

$E_{a'} =$  the energy of the ejectile

$l'=$ the orbital angular momentum of the ejectile

$s'=$ the spin of the ejectile

$j'=$ the total angular momentum of the ejectile 

$\pi_{f}=$ the parity of the ejectile

$\delta_{\pi}(\alpha')=\left\{ \begin{array}{ll}
1 ~~&{\rm if}~~(-1)^{l'}\pi_{f}\Pi_{f}=\Pi \\ 
 0~~& {\rm otherwise}
\end{array}\right.$ 

$\alpha'=$ the designation of channel for the ejectile-residual nucleus final system:

$\alpha'=\{a',\ s',\ E_{a'},\ E_{x},\ I',\ \Pi_f\}$, where $a'$ and  $E_{x}$ are the type of the ejectile and the residual nucleus excitation energy, respectively

$I=$ the spin of target nucleus

$\Pi_{0}=$ the parity of target nucleus

$I'=$ the spin of residual nucleus

$\Pi_f=$ the parity of residual nucleus

$J=$ the total angular momentum of the compound system

$\Pi=$ the parity of the compound system

$D^{comp}=$ the depletion factor so as to take into account for pre-equilibrium and direct effects

$k=$ the wave number of the relative motion

$T=$ the transmission coefficient

$W=$ the correction factor for width fluctuation (WFC).

\begin{figure}[t]
\vspace{0.0cm}
\eject\centerline{\epsfig{file=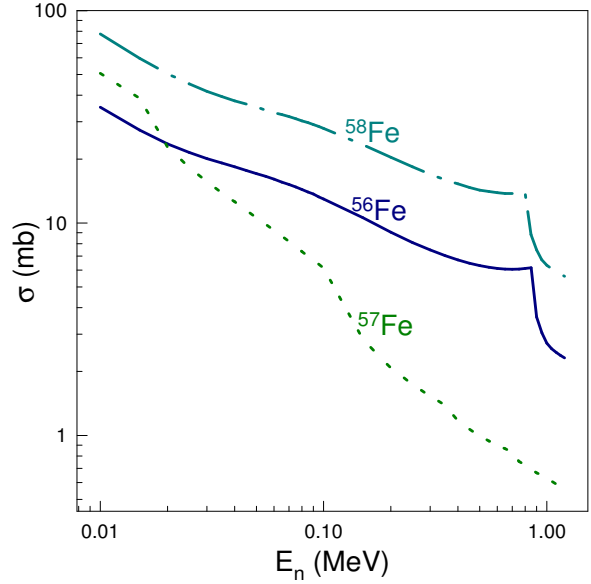,height=7.7cm,width=7.7cm}}
\caption
{Plots of the Hauser-Feshbach estimates of (n,$\gamma$) cross sections as functions of incident neutron energy for Fe isotopes.}
\label{fig1}
\vspace{0.0cm}
\end{figure}
\noindent 

\subsection{ Radiative neutron capture reaction rates }

    The velocities of both the targets and projectiles obey Maxwell- Boltzmann distributions corresponding to ionic plasma temperature $T$ at the site. The astrophysical nuclear reaction rate can be calculated by folding the Maxwell-Boltzmann energy distribution for energies $E$ at the given temperature $T$ with the cross section given by Eq.(2). Additionally, target nuclei exist both in ground and excited states. The relative populations of various energy states of nuclei with excitation energies $E_{x}^{\mu}$ and spins $I^{\mu}$ in thermodynamic equilibrium follows the Maxwell-Boltzmann distribution. In order to distinguish between different excited states the superscript $\mu$ is used along with the incident $\alpha$ channel in the formulas that follow. Taking due account of various target nuclei excited state contributions, the effective nuclear reaction rate in the entrance channel $\alpha\rightarrow\alpha'$ can be finally expressed as 
    
\begin{equation}
 N_{A}\langle\sigma v\rangle_{\alpha\alpha'}^{*}(T)=\left(\frac{8}{\pi m}\right)^{1/2}\frac{N_{A}}{(kT)^{3/2}G(T)}\times\ 
\end{equation}

\begin{center}
   $\displaystyle  \int_{0}^{\infty}\sum_{\mu}\frac{(2I^{\mu}+1)}{(2I^{0}+1)}\sigma_{\alpha\alpha'}^{\mu} (E)E\exp\left(-\frac{E+E_{x}^{\mu}}{kT}\right)dE,$
\end{center}
where $N_{A}$ is the Avogadro number which is equal to 6.023$\times 10^{23}$, $k$ and $m$ are the Boltzmann constant and the reduced mass in the $\alpha$ channel, respectively, and
 
\begin{center}
  $G(T)=\displaystyle \sum_{\mu}(2I^{\mu}+1)/(2I^{0}+1)\exp(-E_{x}^{\mu}/kT)$
\end{center}
is the temperature dependent normalized partition function. By making use of the reciprocity theorem \cite{Ho76}, the reverse reaction cross sections or rates can also be estimated. 

\begin{figure}[t]
\vspace{0.0cm}
\eject\centerline{\epsfig{file=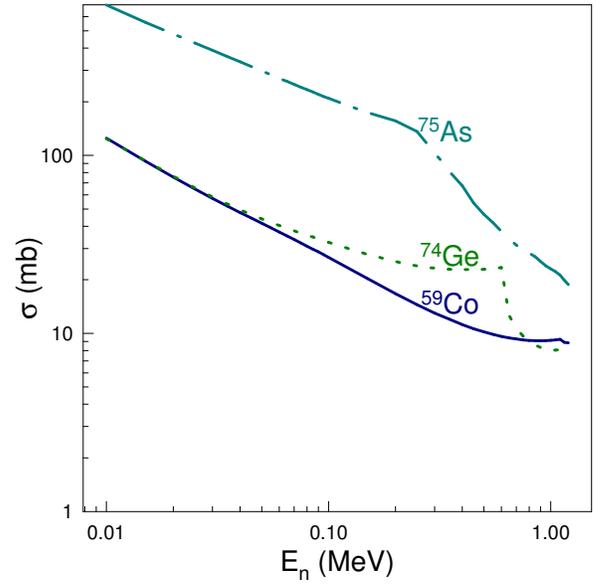,height=7.7cm,width=7.7cm}}
\caption
{Plots of the Hauser-Feshbach estimates of (n,$\gamma$) cross sections as functions of incident neutron energy for $^{59}$Co, $^{74}$Ge and $^{75}$As.}
\label{fig2}
\vspace{0.0cm}
\end{figure}
\noindent

\noindent
\section{ The nuclear physics of r-process }
\label{section3}

    In nuclear astrophysics, the r-process is a series of nuclear reactions which is responsible for the synthesis of approximately half of heavy nuclei resulting origination of elements beyond iron. The neutron star mergers, Type II supernovae and low-mass supernovae \cite{Ba06} are thought to be the three probable r-process candidate sites where the appropriate conditions necessary for nucleosynthesis are expected to exist.

\begin{figure}[t]
\vspace{0.0cm}
\eject\centerline{\epsfig{file=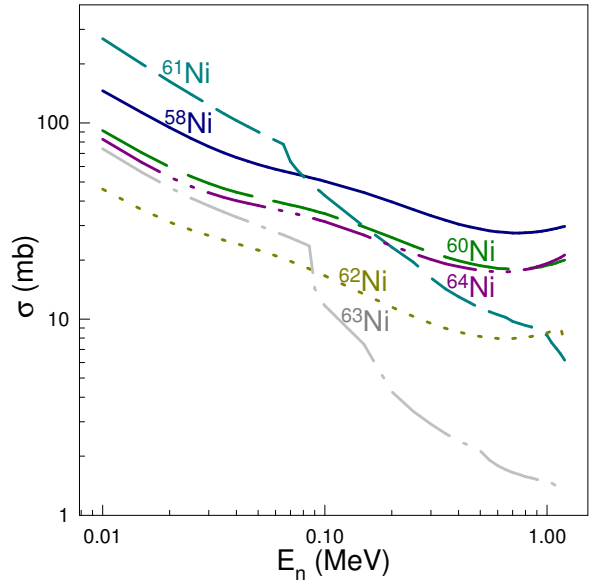,height=7.7cm,width=7.7cm}}
\caption
{Plots of the Hauser-Feshbach estimates of (n,$\gamma$) cross sections as functions of incident neutron energy for Ni isotopes.}
\label{fig3}
\vspace{0.0cm}
\end{figure}
\noindent

    In a Type II supernova, shortly after the extreme compression of electrons $\beta^-$ decay is forbidden. This happens due to the high electron density which occupies all the states available to free electrons up to a Fermi energy that is larger than the $\beta^-$ decay energy. However, these free electrons capture by nuclei do occur causing increasing neutronization of nuclear matter. This fact causes high temperatures and extremely high density $\sim$ 10$^{24}$ of free neutrons per cm$^3$ \cite{Bu57} that can not decay. At this stage it bloats and the expansion cools it and neutron capture by heavy nuclei present in environment advances much rapidly than the $\beta^-$ decay. Consequently, the r-process progresses along the path of the neutron drip line and neutron-rich unstable nuclei are synthesized.

    The dampening process of photo-disintegration, the rapid decrease in the neutron-capture cross section as it approaches closed neutron shells and the magnitude of nuclear stability in the domain of heavy-isotopes are the three major processes which influence the ascent of neutron drip line. Thus the neutron-rich, weakly bound nuclei are formed with neutron separation energies as low as 2 MeV \cite{Bu57,Th04} due to neutron captures during r-process nucleosynthesis. At this stage, the neutron capture is temporarily paused as it reaches the neutron shell closures at N = 50, 82 and 126. These pauses or the so-called waiting points are characterized by increased binding energy compared to heavier isotopes. As a consequence, it results in low neutron capture cross sections and a growth of semi-magic nuclei which are more stable towards $\beta$ decay. Additionally, nuclei beyond the shell closures, owing to their proximity to the drip line, are inclined to  $\beta$ decay briskly. The $\beta$ decay happens before further neutron capture \cite{Wang15} for these nuclei. The waiting point nuclei then prefer $\beta$ decay to move toward stability ahead of any more neutron capture \cite{Bu57}, causing deceleration or freeze-out of the reaction.

    The decrease in the stability of nuclei puts an end to the r-process when its heaviest nuclei develop instability towards spontaneous fission. At this point the total number of nucleons reaches two hundred seventy but before that sufficiently low fission barrier might induce fission at neutron capture and continuing towards neutron drip line \cite{Bo72} terminates. After the decrease in neutron flux, these radioactive nuclei which are highly unstable undergo successive $\beta$ decays rapidly until they approach neutron-rich nuclei which are more stable \cite{Cl68}. In neutron-rich predecessor nuclei, the r-process produces an abundance pattern of radioactive nuclei $\sim$ 10 amu below the peaks of slow neutron-capture process after their decays back to stability whereas the slow neutron-capture process produces an abundance of closed neutron shell stable nuclei.
  
\begin{figure}[t]
\vspace{0.0cm}
\eject\centerline{\epsfig{file=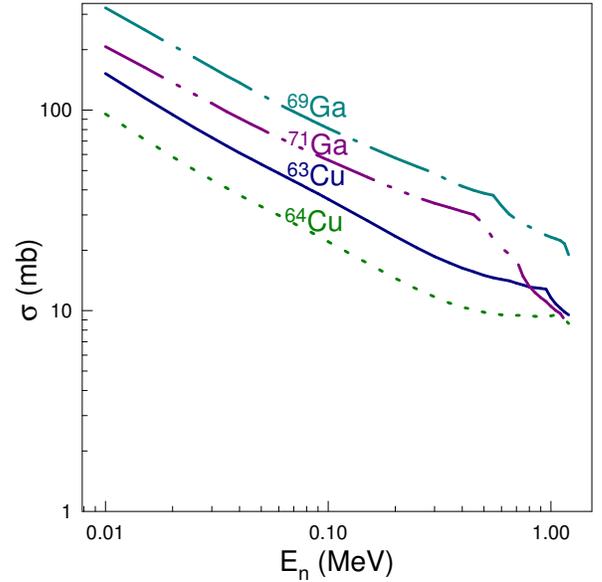,height=7.7cm,width=7.7cm}}
\caption
{Plots of the Hauser-Feshbach estimates of (n,$\gamma$) cross sections as functions of incident neutron energy for Cu and Ga isotopes.}
\label{fig4}
\vspace{0.0cm}
\end{figure}
\noindent 
\clearpage  

\begin{figure}[ht!]
\vspace{0.0cm}
\eject\centerline{\epsfig{file=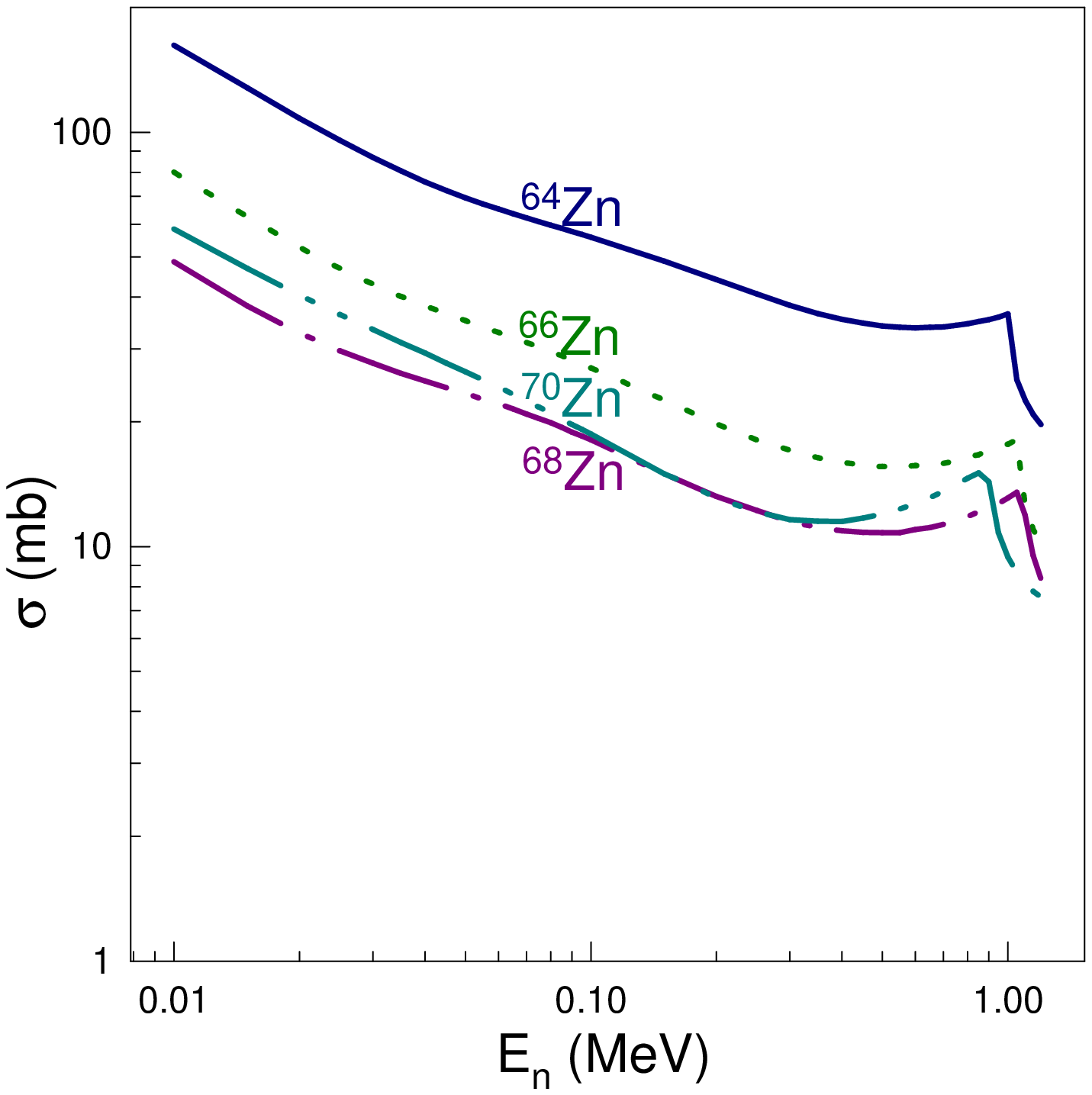,height=7.7cm,width=7.7cm}}
\caption
{Plots of the Hauser-Feshbach estimates of (n,$\gamma$) cross sections as functions of incident neutron energy for Zn isotopes.}
\label{fig5}
\vspace{0.0cm}
\end{figure}
\noindent 
    
\begin{figure}[b]
\vspace{0.0cm}
\eject\centerline{\epsfig{file=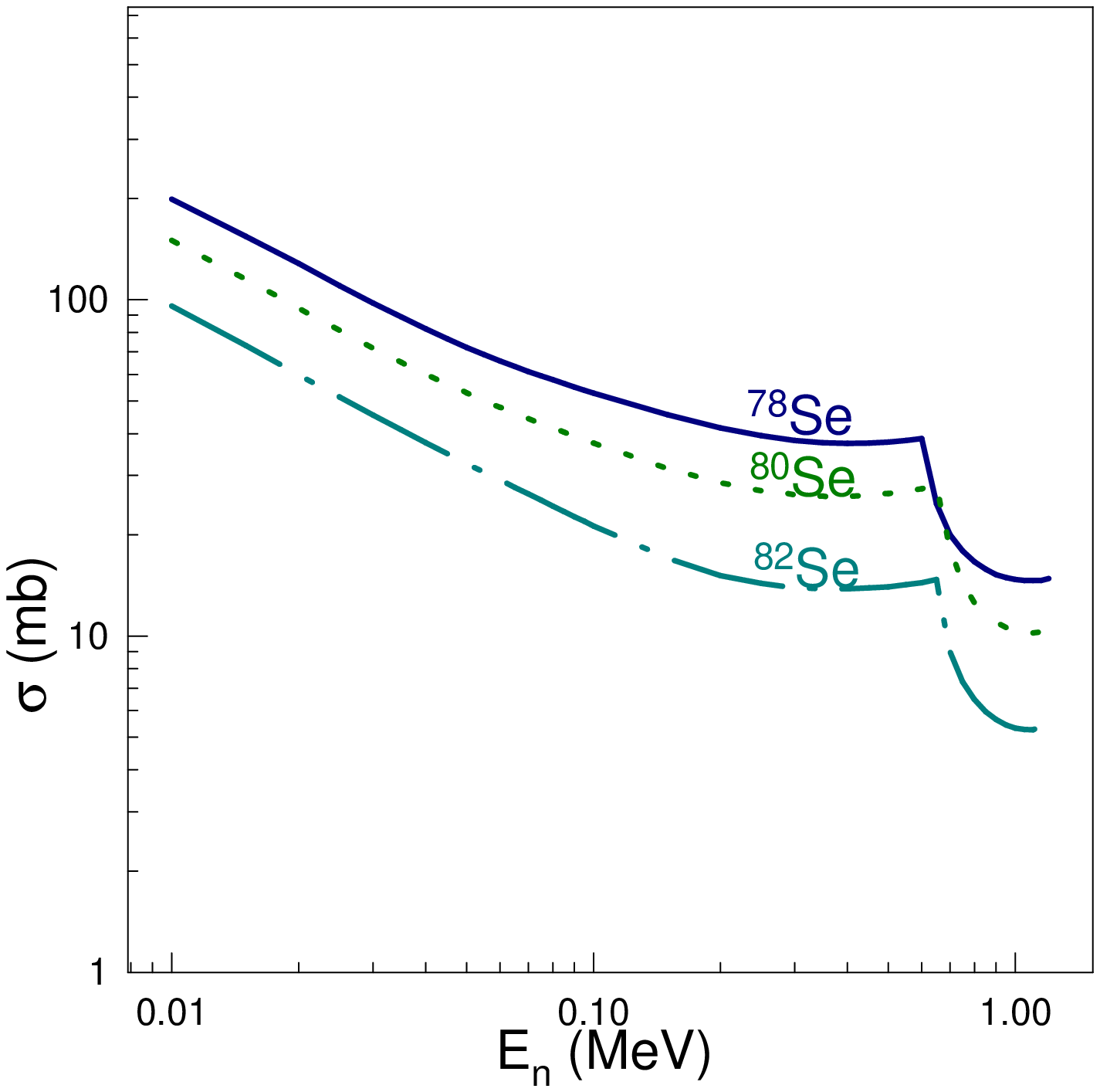,height=7.7cm,width=7.7cm}}
\caption
{Plots of the Hauser-Feshbach estimates of (n,$\gamma$) cross sections as functions of incident neutron energy for Se isotopes.}
\label{fig6}
\vspace{0.0cm}
\end{figure}
\noindent  

\begin{figure}[ht!]
\vspace{0.0cm}
\eject\centerline{\epsfig{file=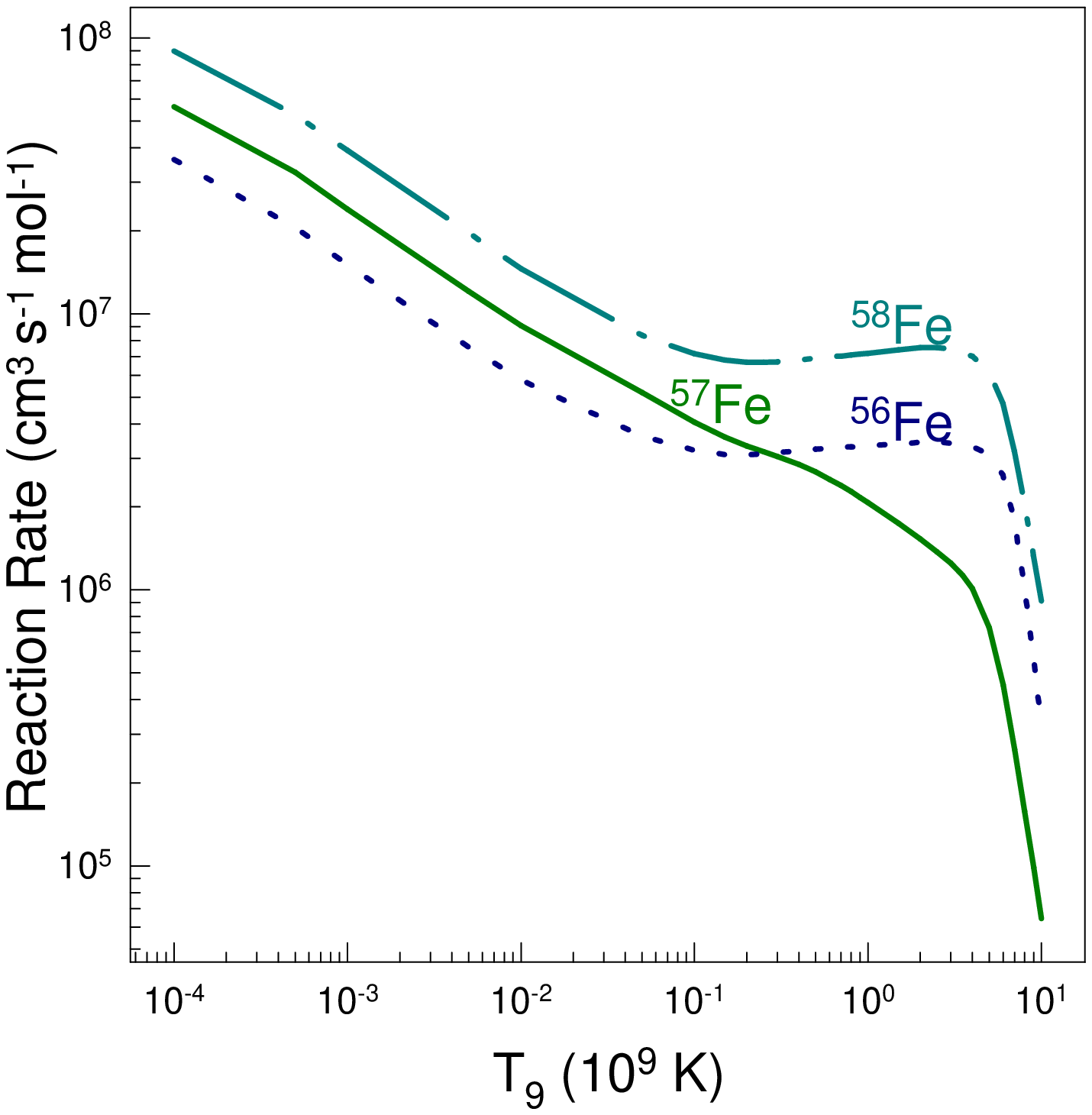,height=7.7cm,width=7.7cm}}
\caption
{Plots of the (n,$\gamma$) reaction rates as functions of T$_9$ for Fe isotopes.}
\label{fig7}
\vspace{0.0cm}
\end{figure}
\noindent  

\begin{figure}[b]
\vspace{0.0cm}
\eject\centerline{\epsfig{file=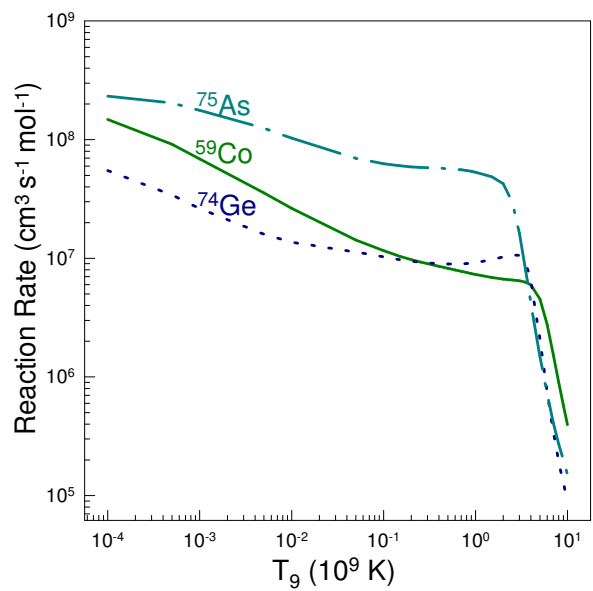,height=7.7cm,width=7.7cm}}
\caption
{Plots of the (n,$\gamma$) reaction rates as functions of T$_9$ for $^{59}$Co, $^{74}$Ge and $^{75}$As.}
\label{fig8}
\vspace{0.0cm}
\end{figure}
\noindent  
\clearpage

\begin{figure}[ht!]
\vspace{0.0cm}
\eject\centerline{\epsfig{file=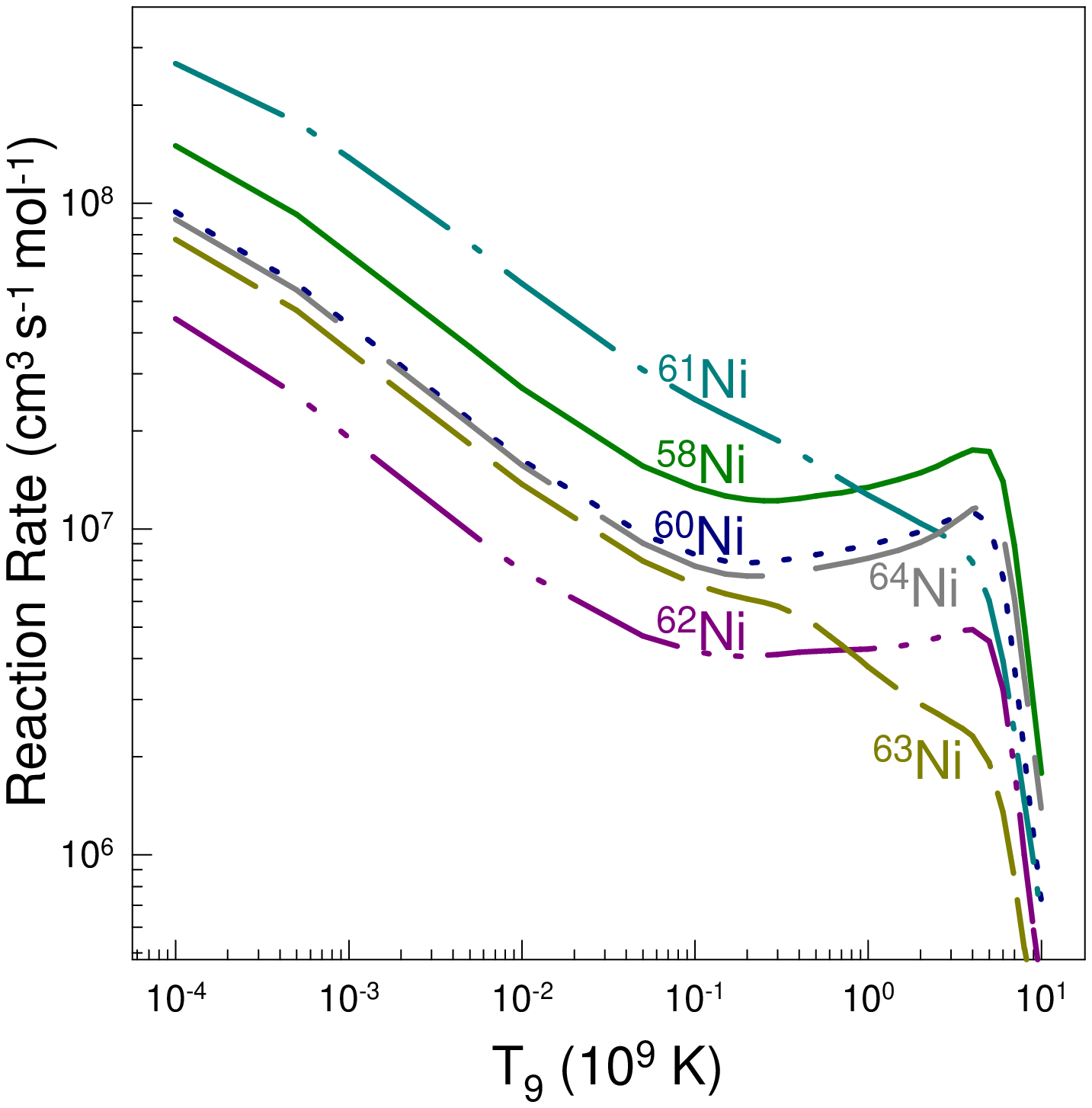,height=7.7cm,width=7.7cm}}
\caption
{Plots of the (n,$\gamma$) reaction rates as functions of T$_9$ for Ni isotopes.}
\label{fig9}
\vspace{0.0cm}
\end{figure}
\noindent  

\begin{figure}[b]
\vspace{0.0cm}
\eject\centerline{\epsfig{file=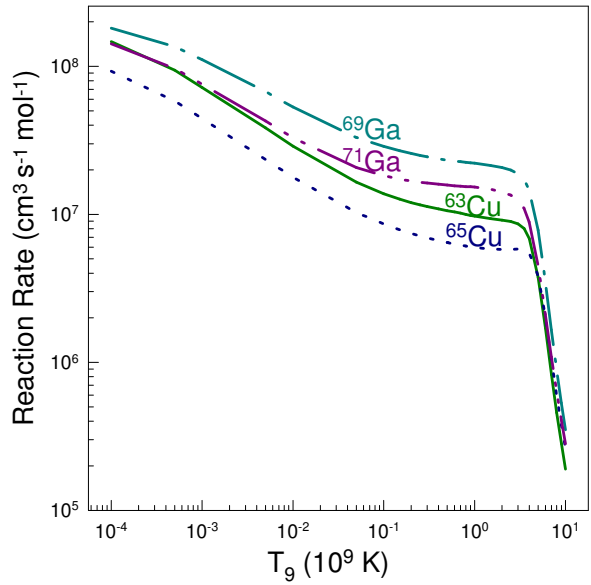,height=7.7cm,width=7.7cm}}
\caption
{Plots of the (n,$\gamma$) reaction rates as functions of T$_9$ for Cu and Ga isotopes.}
\label{fig10}
\vspace{0.0cm}
\end{figure}
\noindent  

\begin{figure}[ht!]
\vspace{0.0cm}
\eject\centerline{\epsfig{file=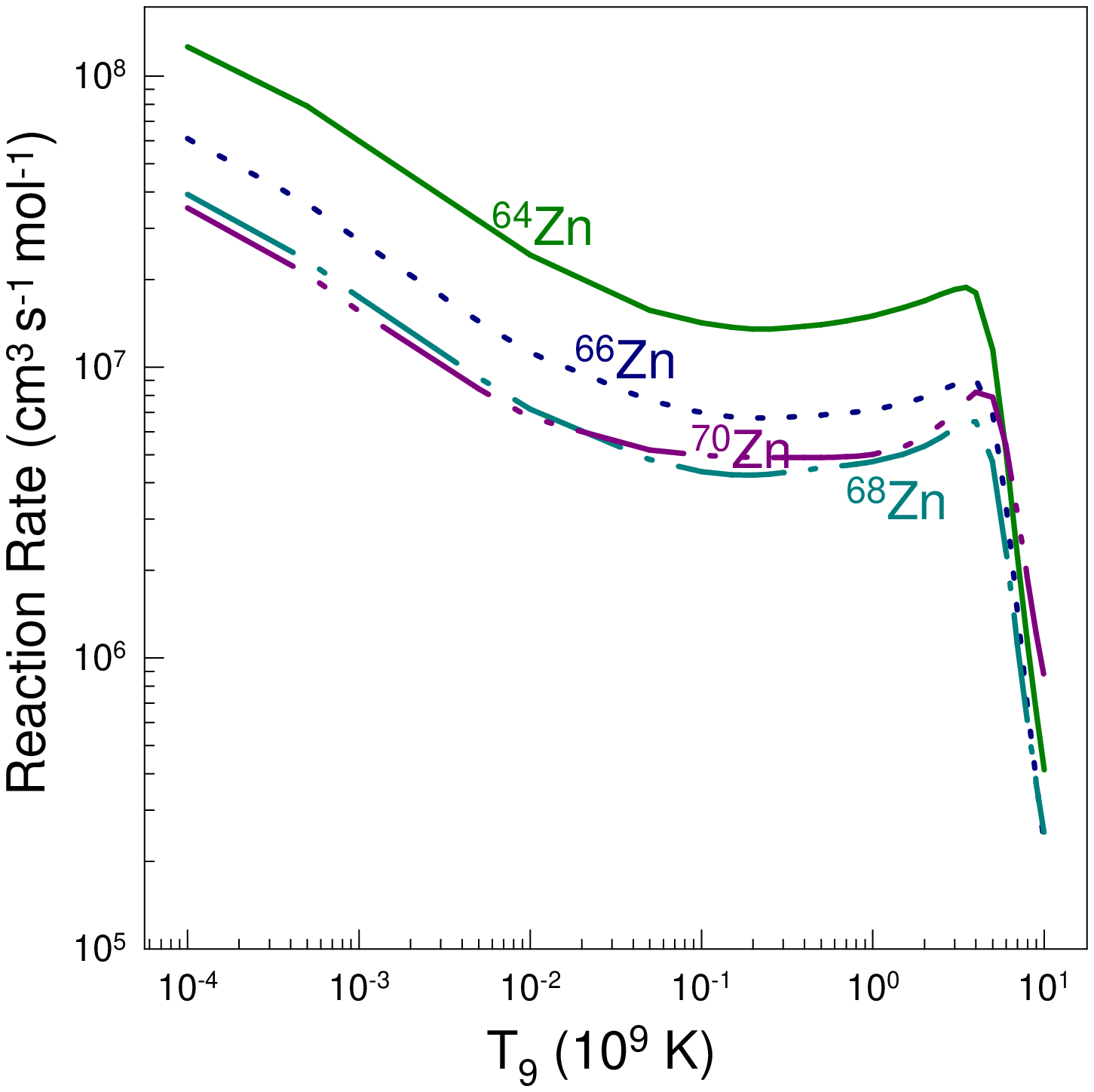,height=7.7cm,width=7.7cm}}
\caption
{Plots of the (n,$\gamma$) reaction rates as functions of T$_9$ for Zn isotopes.}
\label{fig11}
\vspace{0.0cm}
\end{figure}
\noindent  

\begin{figure}[b]
\vspace{0.0cm}
\eject\centerline{\epsfig{file=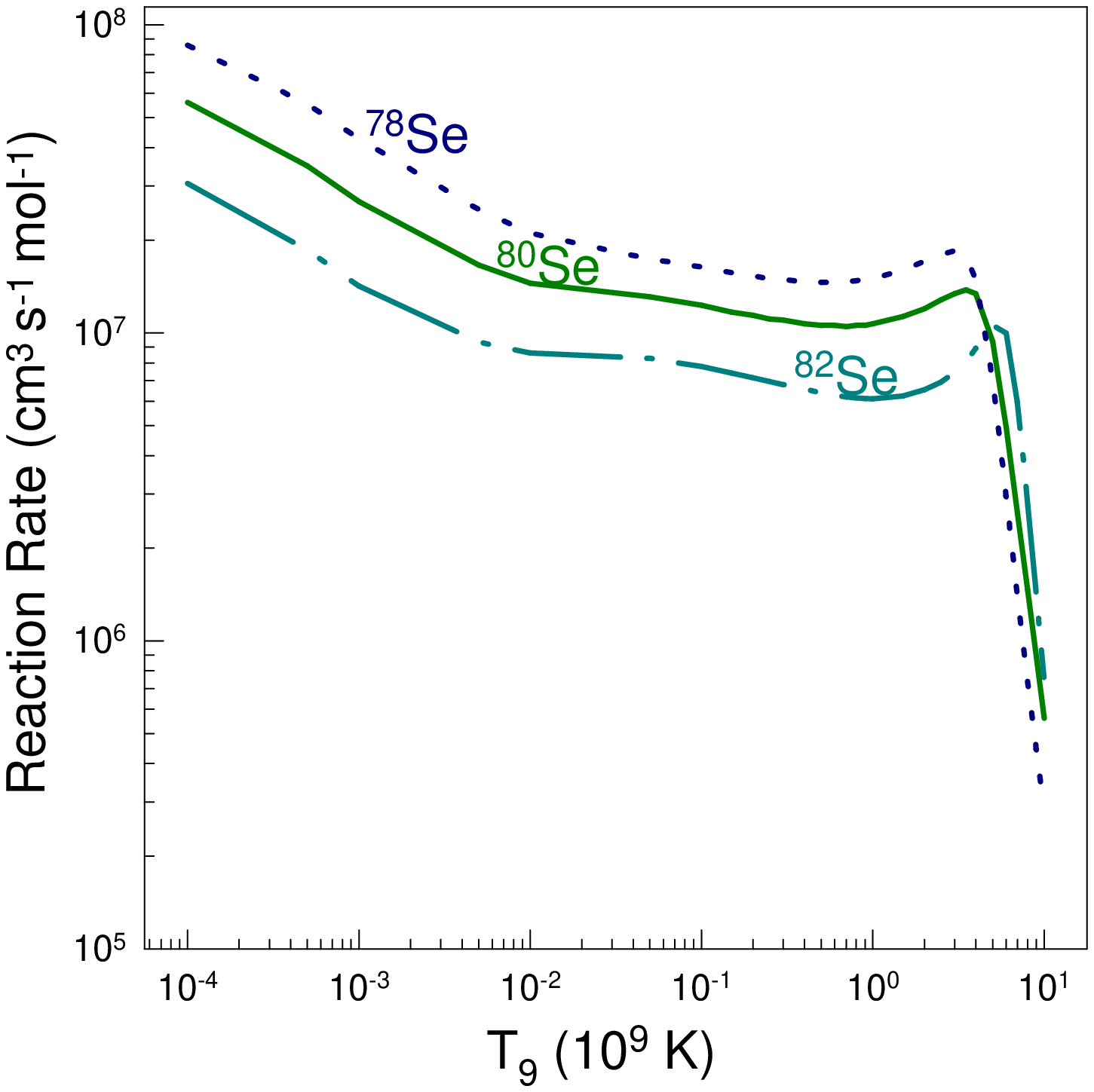,height=7.7cm,width=7.7cm}}
\caption
{Plots of the (n,$\gamma$) reaction rates as functions of T$_9$ for Se isotopes.}
\label{fig12}
\vspace{0.0cm}
\end{figure}
\noindent  
\clearpage

\begin{figure}[ht!]
\vspace{0.0cm}
\eject\centerline{\epsfig{file=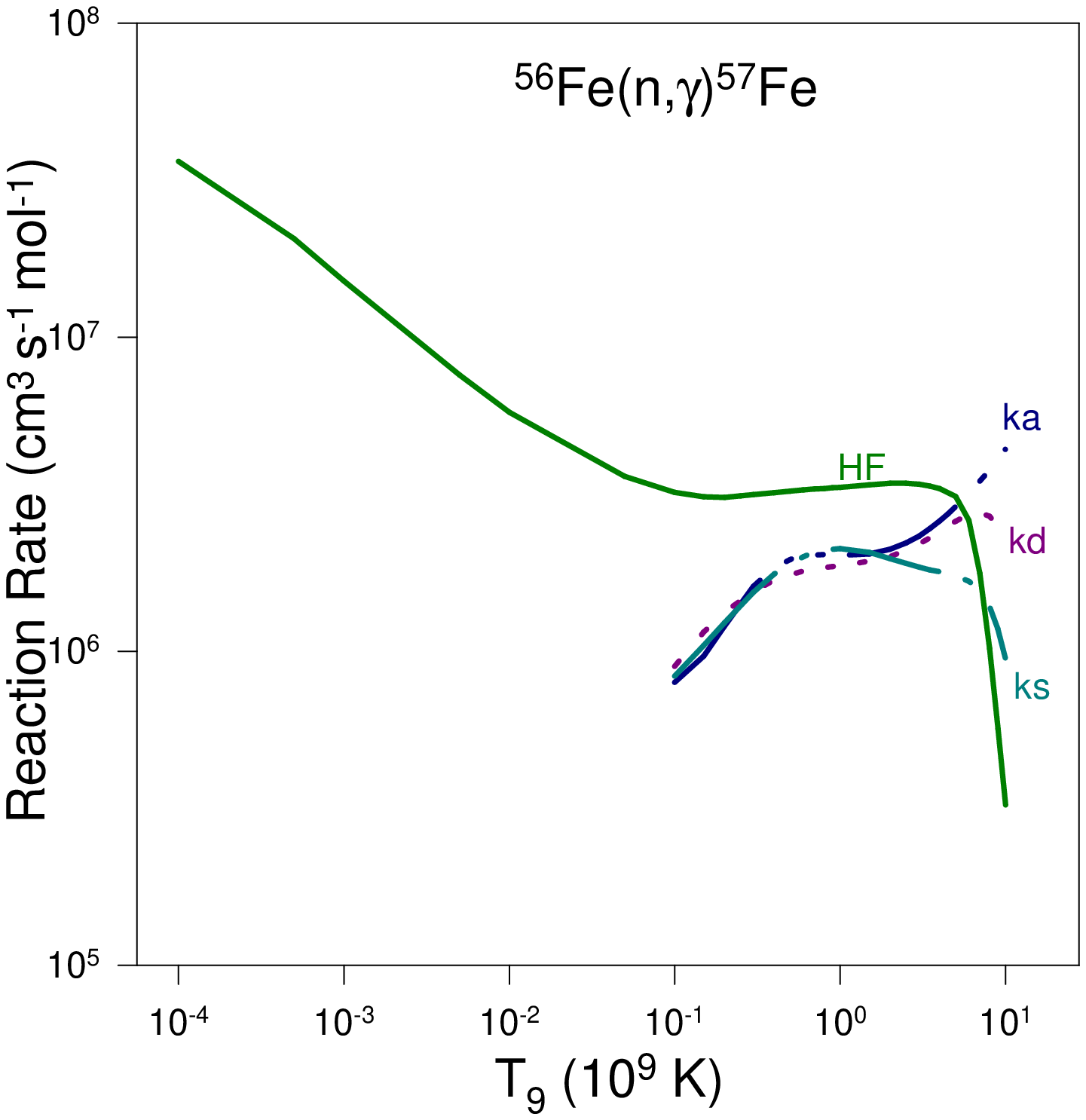,height=7.7cm,width=7.7cm}}
\caption
{Comparison of the predictions of $^{56}$Fe(n,$\gamma$) reaction rates (HF) with the data from JINA REACLIB \cite{kakd,ks}.}
\label{fig13}
\vspace{0.0cm}
\end{figure}
\noindent  

\begin{figure}[b]
\vspace{0.0cm}
\eject\centerline{\epsfig{file=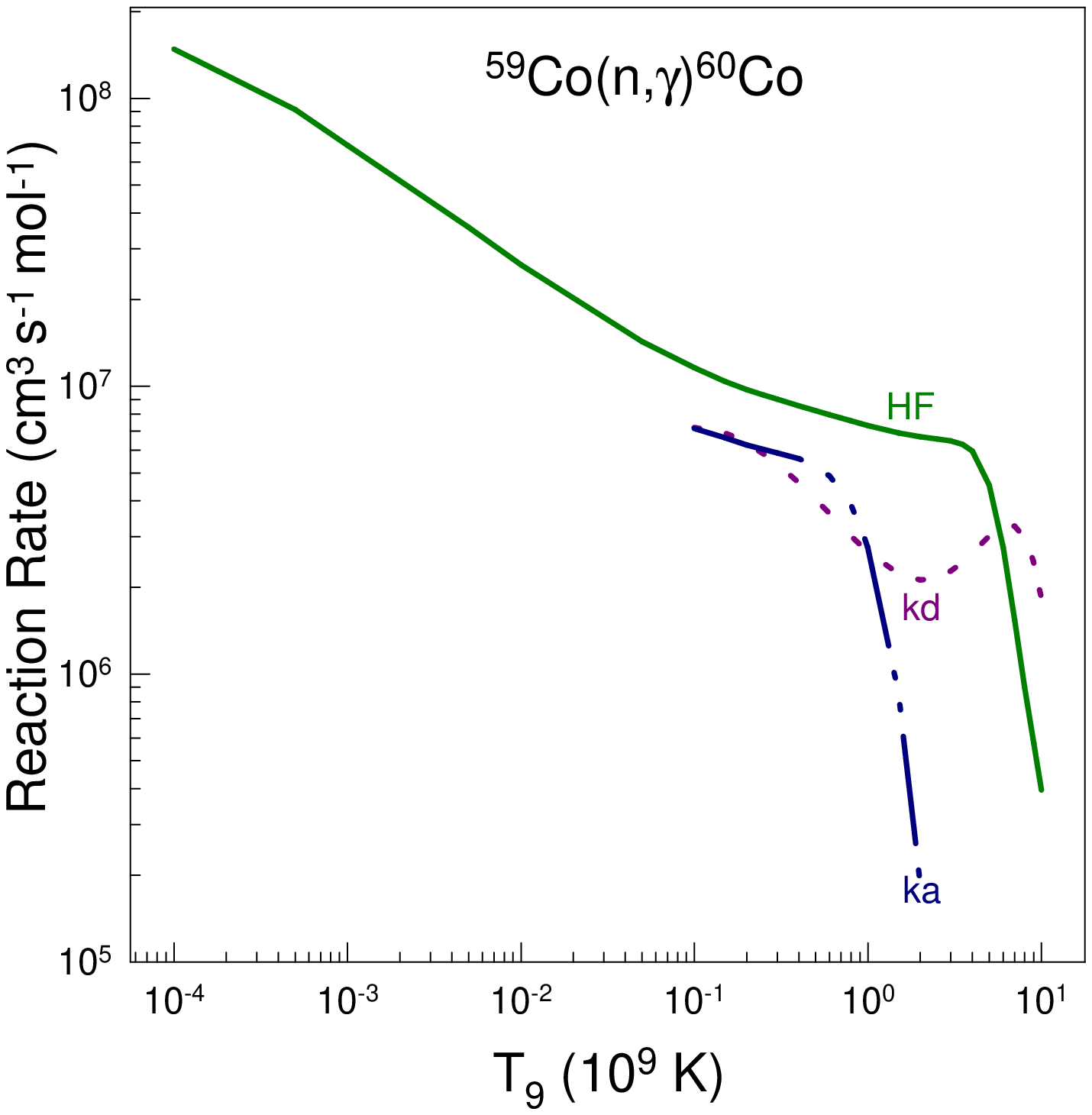,height=7.7cm,width=7.7cm}}
\caption
{Comparison of the predictions of $^{59}$Co(n,$\gamma$) reaction rates (HF) with the data from JINA REACLIB \cite{kakd,ks}.}
\label{fig14}
\vspace{0.0cm}
\end{figure}
\noindent  

\begin{figure}[ht!]
\vspace{0.0cm}
\eject\centerline{\epsfig{file=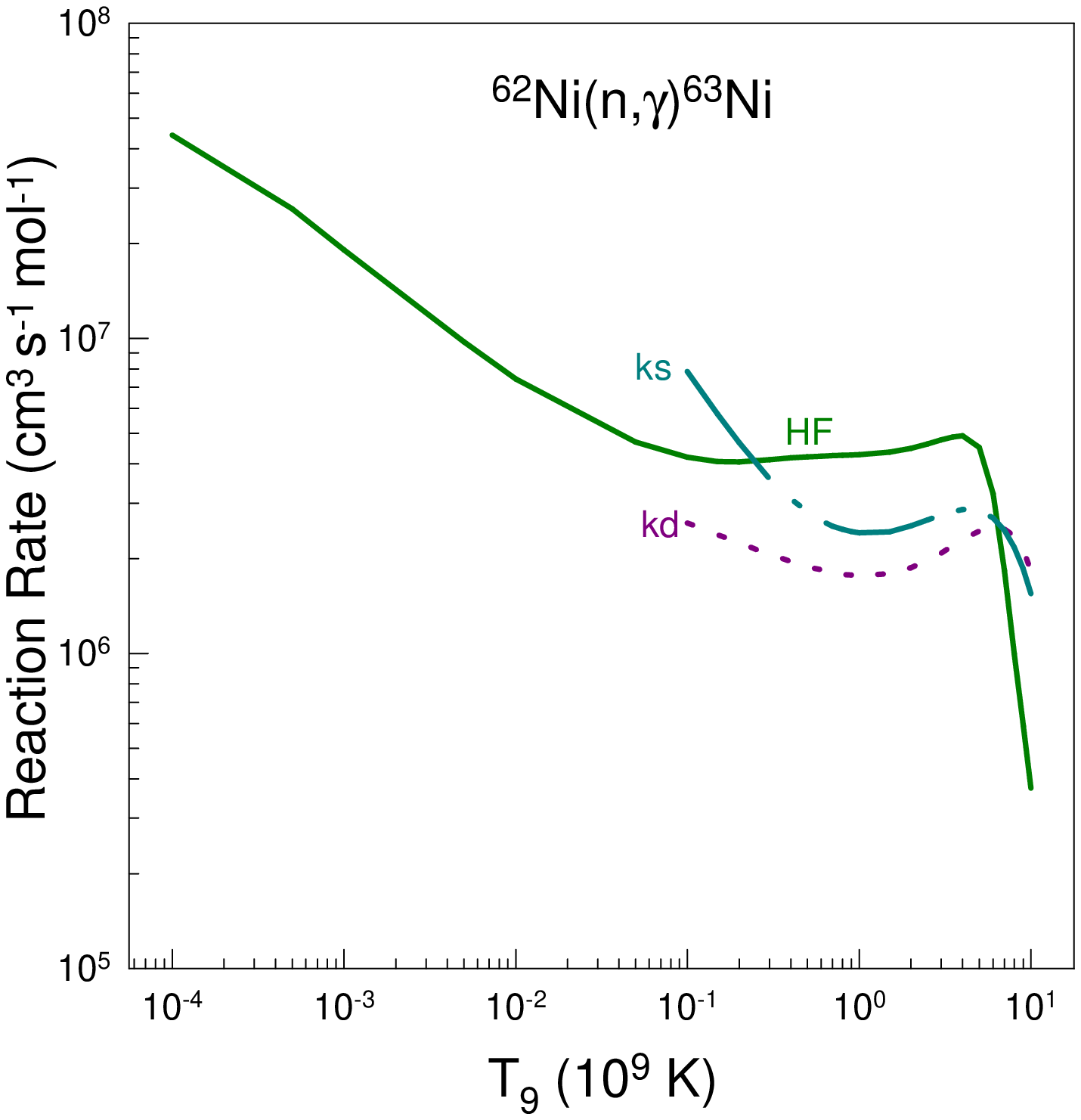,height=7.7cm,width=7.7cm}}
\caption
{Comparison of the predictions of $^{62}$Ni(n,$\gamma$) reaction rates (HF) with the data from JINA REACLIB \cite{kakd,ks}.}
\label{fig15}
\vspace{0.0cm}
\end{figure}
\noindent  

\begin{figure}[b]
\vspace{0.0cm}
\eject\centerline{\epsfig{file=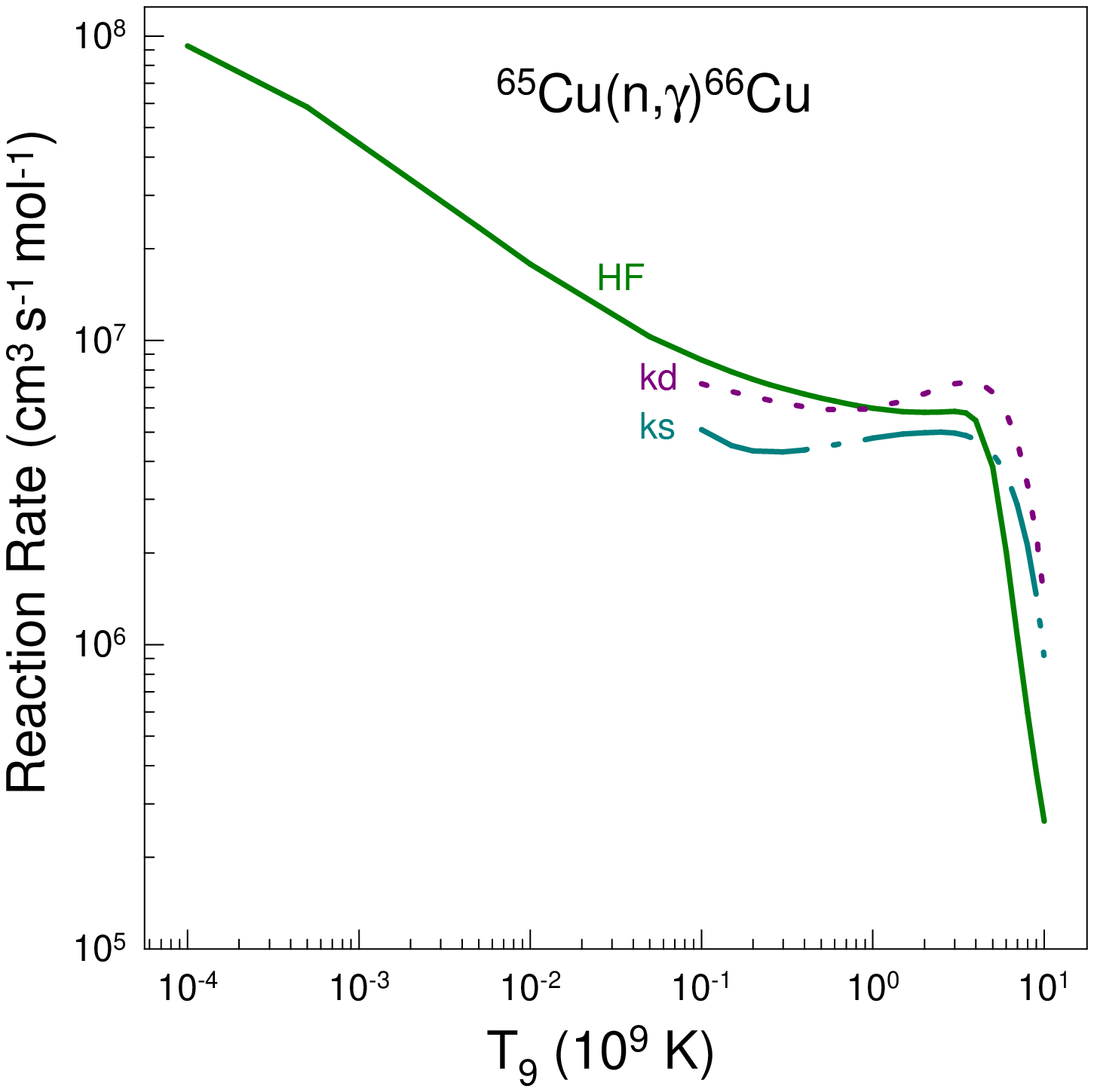,height=7.7cm,width=7.7cm}}
\caption
{Comparison of the predictions of $^{65}$Cu(n,$\gamma$) reaction rates (HF) with the data from JINA REACLIB \cite{kakd,ks}.}
\label{fig16}
\vspace{0.0cm}
\end{figure}
\noindent  
\clearpage

\begin{figure}[ht!]
\vspace{0.0cm}
\eject\centerline{\epsfig{file=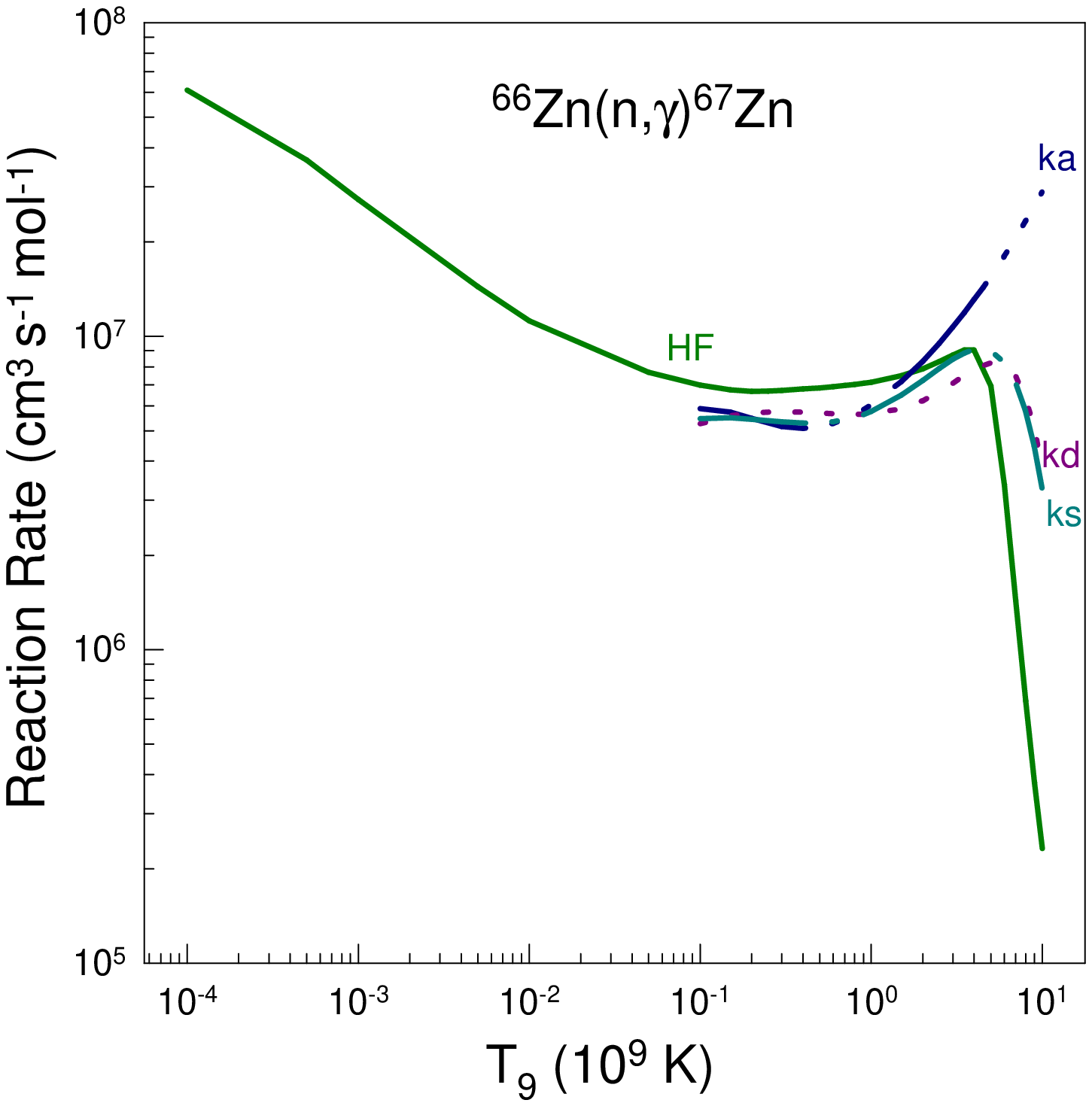,height=7.7cm,width=7.7cm}}
\caption
{Comparison of the predictions of $^{66}$Zn(n,$\gamma$) reaction rates (HF) with the data from JINA REACLIB \cite{kakd,ks}.}
\label{fig17}
\vspace{0.0cm}
\end{figure}
\noindent  

\begin{figure}[b]
\vspace{0.0cm}
\eject\centerline{\epsfig{file=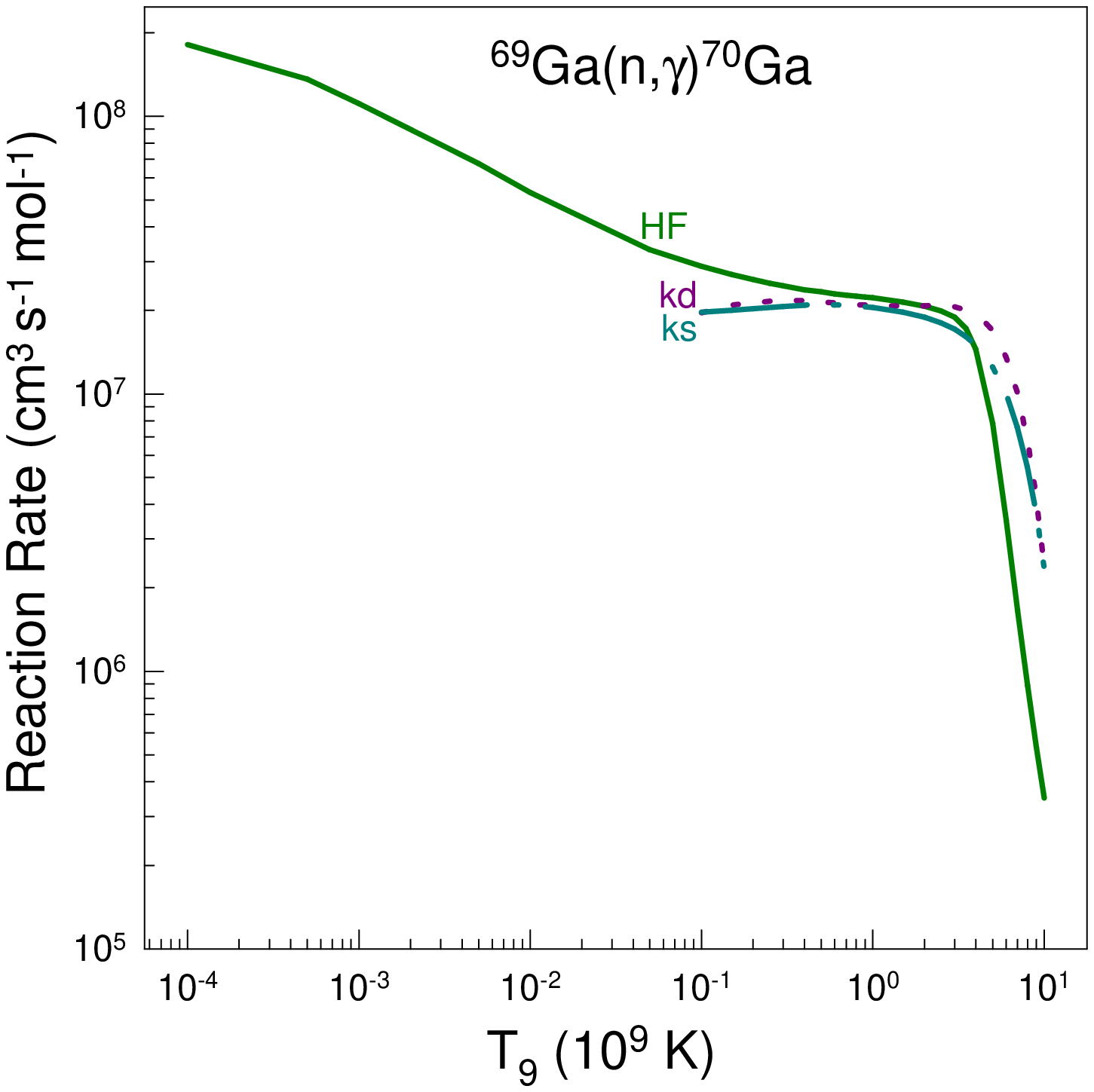,height=7.7cm,width=7.7cm}}
\caption
{Comparison of the predictions of $^{69}$Ga(n,$\gamma$) reaction rates (HF) with the data from JINA REACLIB \cite{kakd,ks}.}
\label{fig18}
\vspace{0.0cm}
\end{figure}
\noindent  

\begin{figure}[ht!]
\vspace{0.0cm}
\eject\centerline{\epsfig{file=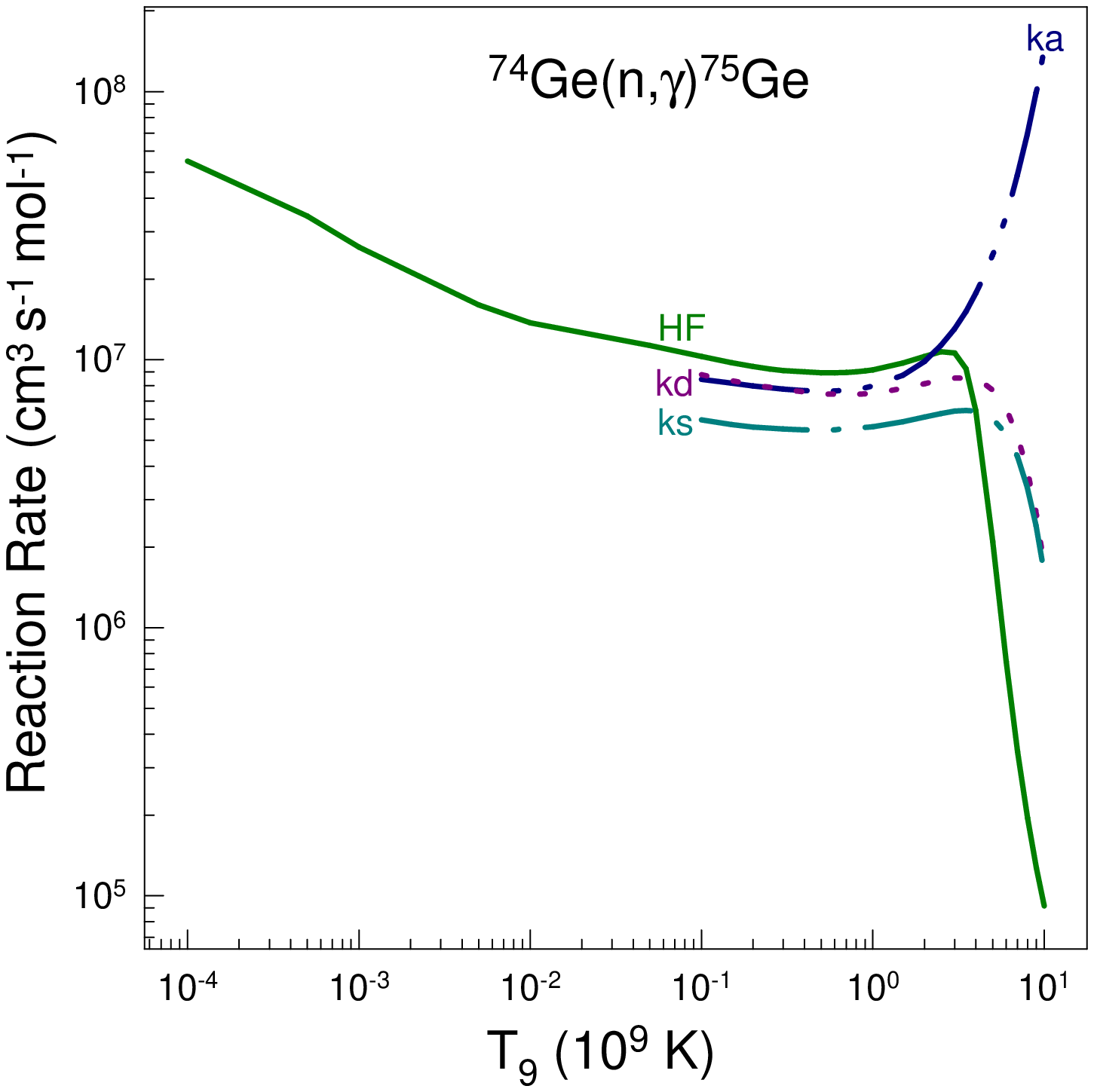,height=7.7cm,width=7.7cm}}
\caption
{Comparison of the predictions of $^{74}$Ge(n,$\gamma$) reaction rates (HF) with the data from JINA REACLIB \cite{kakd,ks}.}
\label{fig19}
\vspace{0.0cm}
\end{figure}
\noindent  

\begin{figure}[b]
\vspace{0.0cm}
\eject\centerline{\epsfig{file=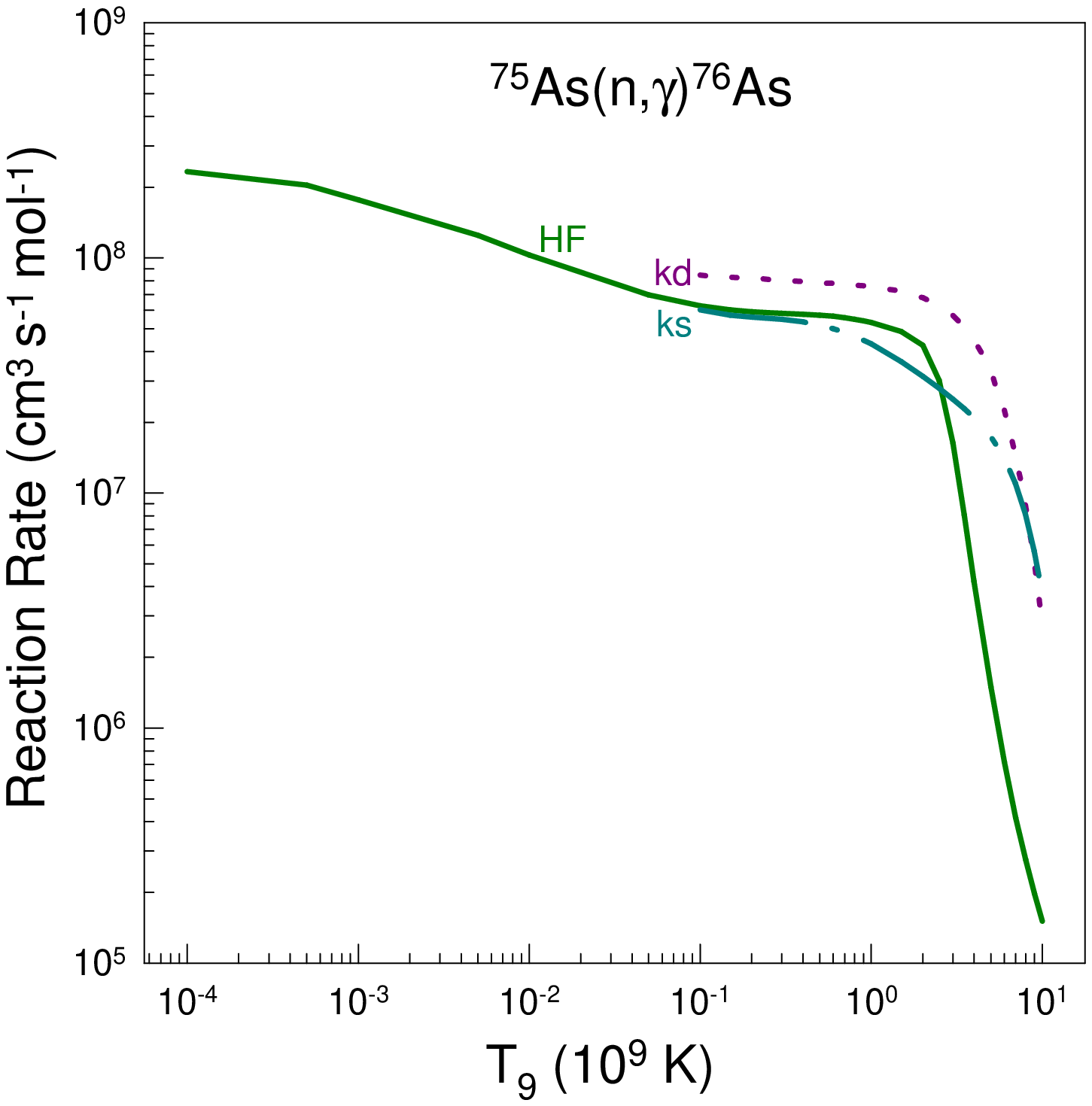,height=7.7cm,width=7.7cm}}
\caption
{Comparison of the predictions of $^{75}$As(n,$\gamma$) reaction rates (HF) with the data from JINA REACLIB \cite{kakd,ks}.}
\label{fig20}
\vspace{0.0cm}
\end{figure}
\noindent  
\clearpage

\begin{figure}[htbp]
\vspace{0.0cm}
\eject\centerline{\epsfig{file=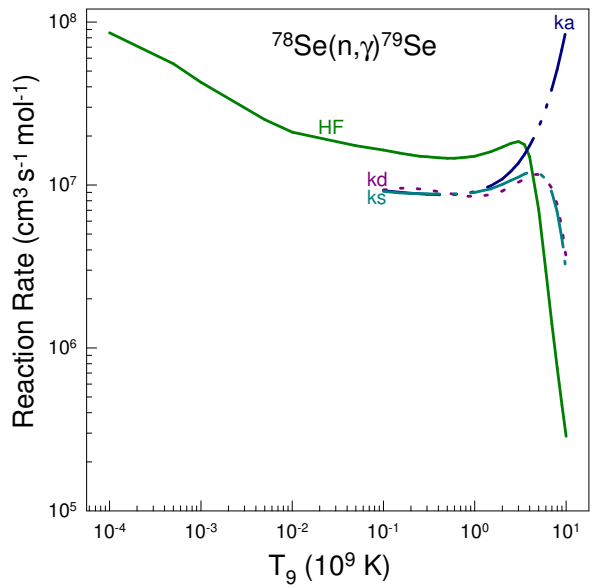,height=7.7cm,width=7.7cm}}
\caption
{Comparison of the predictions of $^{78}$Se(n,$\gamma$) reaction rates (HF) with the data from JINA REACLIB \cite{kakd,ks}.}
\label{fig21}
\vspace{0.0cm}
\end{figure}
\noindent  

\noindent
\section{ Results and discussion }
\label{section4}

\subsection{ Radiative neutron capture cross section calculations  }

    We have calculated theoretically the radiative neutron capture (n,$\gamma$) cross sections of astrophysical importance for Fe, Co, Ni, Cu, Zn, Ga, Ge, As and Se isotopes using the Hauser-Feshbach statistical model reaction calculations. The calculations have been performed using the most recent level density based on temperature-dependent Hartree-Fock-Bogolyubov calculations using the Gogny force and for the gamma-ray strength function Brink-Axel Lorentzian has been used \cite{Talys}. In Figs.1-6, the plots of these estimates for the radiative neutron capture cross sections as functions of incident neutron energy for the above mentioned elements and their different isotopes have been illustrated.

\subsection{ Reaction rate calculations for neutron rich nuclei }
 
    The cross sections for nuclear reaction and its convolution with Maxwell-Boltzmann distribution of energies are important for the explanation of various processes occurring under extreme conditions \cite{Bu57,Fo64,Cl83}. In the main-sequence stars and compact stars which are in their ultimate stages of evolution, such environments of very high density and temperature prevail. The exothermic fusion reactions causes nuclear explosions in the surface layers of the accreting white dwarfs (nova events), in the cores of massive accreting white dwarfs (type Ia supernovae) \cite{Ni97,Ho06} and in the surface layers of accreting neutron stars (type I X-ray bursts and superbursts \cite{St06,Sc03,Cu06,Gu07}). Precise knowledge of the rates of thermonuclear reactions obtained by folding Maxwell-Boltzmann distribution of energies with energy dependent cross sections becomes necessary for describing these astrophysical phenomena. The Maxwellian-averaged thermonuclear reaction rate per particle pair $<\sigma v>$ at temperature $T$, can be represented by the integral \cite{Ad11,Fo67,Bo08} described below:

\begin{equation}
 <\sigma v> = \Big[\frac{8}{\pi m (k T)^3 } \Big]^{1/2} \int \sigma(E) E \exp(-E/k T) dE,
\label{seqn3}
\end{equation}
\noindent
where $v$ is the relative velocity, $E$ is the energy in the centre-of-mass system, $k$ and $m$ are the Boltzmann constant and the reduced mass of the reacting nuclei, respectively. Thus, the reaction rate between two nuclei can be written as $r_{12}=\frac{n_1n_2}{1+\delta_{12}}<\sigma v>$ where $n_1$ and $n_2$ are the number densities of nuclei of types 1 and 2. The Kronecker delta $\delta_{12}$ prevents double counting in case of identical particles. In Figs.7-12, the plots of the (n,$\gamma$) reaction rates as functions of T$_9$ for Fe, Co, Ni, Cu, Zn, Ga, Ge, As and Se isotopes have been shown.

    Some key reactions \cite{Su14} which may have particularly large impact on the final abundances in the A $\sim$ 80 region have been investigated. Several works \cite{Fo88,Ma89,Sm93,An99,An04} regarding reaction rates have been done in past. Except a few neutron induced reactions, all other reaction rates have temperature dependencies.  The neutron capture cross sections show $\approx 1/v$ behavior at very low energies in the thermal domain. Therefore in Eq.(3), using $\sigma(E) \propto E^{-1/2}$ one finds immediately that at thermal energies the reaction rates are more or less constant with respect to temperature. However, for very low energy neutrons ($\sim$ 0.025 eV) only this fact is true where energies are $\sim$ eV and below. On the other hand, at energies in the region of astrophysical interest, the capture cross sections for the neutron induced reactions can be best described by $\sigma(E)=\frac{R(E)}{v}$ \cite{Bl55}, where $R(E)$ varies gently as a function of energy \cite{Mu10}. It is, therefore, similar to the astrophysical S-factor and one expects $\langle\sigma v\rangle$ to be temperature dependent. Due to the above reason, since Bao-K\"appeler fit \cite{baka} to experimental has been independent of temperature we have excluded it and retained I. Dillman et al. (ka,kd) \cite{kakd} and KADoNiS (ks) \cite{ks}  for comparison with experimental results. The comparison of the present (n,$\gamma$) reaction rate calculations with the JINA REACLIB reaction rates \cite{baka,kakd,ks} has been presented in Figs.13-21. 
    
\noindent
\section{ Summary and conclusion }
\label{section5}

        To summarize, in the present work the theoretical predictions of radiative neutron capture (n,$\gamma$) cross sections of astrophysical importance and the reaction rates for Fe, Co, Ni, Cu, Zn, Ga, Ge, As and Se isotopes using the Hauser-Feshbach statistical model reaction calculations have been investigated. It is observed that the experimental results are uncertain by a few orders of magnitude for nuclei even in the vicinity of the valley of stability. Some key reactions \cite{Su14} which may have significantly large impact on the final abundances in the region of mass number around eighty have been explored. The calculations of the (n,$\gamma$) reaction rates have been compared with the JINA REACLIB reaction rates. Since in several cases large deviations among fits to experimental data of ka, kd and ks do exist, estimates of present calculations can be termed as good. In addition, it is recognized that the uncertainties due to the factors such as level densities and mass models may have substantial effects on the rates while the low-energy upbend in the $\gamma$-strength function has a little (though non-negligible) effect on the rates.

    To conclude, it is envisaged that to constrain the (n,$\gamma$) reaction rates near the mass region eighty there is an acute need of more data. In order to exclude or establish certain model inputs, new experimental techniques, namely, the surrogate method for neutron rich nuclei \cite{Koz12} and the beta-Oslo method \cite{Sp14} may contribute some crucial information of paramount importance.
     
\begin{acknowledgements}

    One of the authors (DNB) acknowledges support from Science and Engineering Research Board, Department of Science and Technology, Government of India, through Grant No. CRG/2021/007333.

\end{acknowledgements}

\end{document}